\documentclass[12pt]{article}
\usepackage{epsf,latexsym}
\usepackage{amsmath,amsfonts,amssymb,amsthm,cite}
\usepackage{graphicx}

\theoremstyle{definition}                    

\theoremstyle{remark}

\numberwithin{equation}{section}             

\usepackage[ps,dvips,arc,frame]{xy}

\newcommand{\rxy}[1]{{\begin{xy}
0;<2mm,0mm>:<0mm,2mm>::0;0,#1\end{xy}}}

\epsfverbosetrue
\newcommand{\bb}{\begin{eqnarray}}
\newcommand{\ee}{\end{eqnarray}}
\newcommand{\eee}{\nonumber\end{eqnarray}}
\newcommand{\pp}[1]{\begin{pmatrix} #1 \end{pmatrix}}

\newcommand{\rxyz}[2]{{\begin{xy} 0;<2mm,0mm>:<0mm,2mm>::0;0,
,(5,-5)*\cir(#1,0){} ,(10,-5)*\cir(#1,0){} ,(15,-5)*\cir(#1,0){}
,(5,-10)*\cir(#1,0){} ,(10,-10)*\cir(#1,0){} ,(15,-10)*\cir(#1,0){}
,(5,-15)*\cir(#1,0){} ,(10,-15)*\cir(#1,0){} ,(15,-15)*\cir(#1,0){}
#2\end{xy}}}

\begin{document}

\thispagestyle{empty}

\begin{center}
INSTITUT F\"UR THEORETISCHE PHYSIK UND ASTROPHYSIK\\
Christian-Albrechts-Universit\"at zu Kiel\footnote{\, Leibnizstra\ss e 15, D-24098 Kiel, Germany}
\end{center}

\vspace{1cm}

\begin{center}
{\Large\textbf{Finding the Standard Model of
Particle Physics \\[0.2cm] A Combinatorial Problem}} 
\end{center}

\begin{center}
{\large Jan-Hendrik Jureit
\footnote{\, Universit\'e de Provence}${}^,$
\footnote{\, jureit@cpt.univ-mrs.fr}${}^,$ ${}^5$
\quad Christoph A. Stephan$\,\,^{2,}$
\footnote{\, stephan@cpt.univ-mrs.fr}${}^,$ 
\footnote{\, Centre de Physique Th\'eorique\, Unit\'e Mixed de
Recherche (UMR) 6207 du CNRS \\  \indent \, \quad et des Universit\'es Aix-Marseille 1 et 2  Sud Toulon-Var, Laboratoire \\ \indent \, \quad affili\'e \`a la 
FRUMAM (FR 2291)}}

\vspace{2cm}

{\large\textbf{Abstract}}
\end{center} 

We present a combinatorial problem which consists in finding irreducible Krajewski diagrams
from finite geometries. 
This problem boils down to placing arrows into a quadratic array with some additional constrains.
The Krajewski diagrams play a central role in the
description of finite noncommutative geometries. They allow to localise the standard model of particle
physics within the set of all Yang-Mills-Higgs models.

\vspace{1cm}

\vskip 1truecm

\vskip 1truecm

\noindent March 2005
\vskip 1truecm

\newpage

\section{Introduction}
In this paper we present an algebraic problem which has its roots in noncommutative geometry.
Given the set of all square matrices $M \in M_n(\mathbb{Z})$, with integer entries and for fixed $n$. These matrices are called
multiplicity matrices. Define a partial order in
$M_n(\mathbb{Z})$ by $M \geq M'$ when $M_{ij}$ and $M_{ij}'$ have the same sign and
$|M _{ij}|\geq|M_{ij}'|$ for all $i,j=1,\ldots, n$.
The task is now to find all the minimal multiplicity matrices which obey two further conditions.
First, the determinant of $M$ has to be non-zero. The second condition asserts that
for any non vanishing matrix element there has to exist a second element of oposite sign in the same row
or in the same column. It is straight forward to translate this problem into a combinatorial problem
using diagramtic
language with arrows as basic elements.
We will give a simple set of rules how to fit arrows into a quadratic diagram body. The resulting
diagrams are called Krajewski diagrams.  
They can be reduced by combining or ``clipping'' the arrows in the diagram in accordance with
the rules. To each Krajewski diagram a  multiplicity matrix will be associated. 
Our aim will be to find those Krajewski diagrams with a minimal number of arrows that are also as much reduced as possible while the determinant of the multiplicity matrix
is non-zero. These diagrams will be called irreducible.

The rules to construct Krajewski diagrams follow directly from Alain Connes non-commutative geometry
\cite{book}. They are due to a special class of noncommutative geometries called almost-commutative
geometries. These unify the general theory of relativity with the classical field theory of the
standard model of particle physics \cite{cc}. It is possible to give a complete classification of 
almost-commutative geometries \cite{Kraj} which can be narrowed under addition of some physical assumptions
to a classification of a class of Yang-Mills-Higgs theories.  Within these theories the standard model of particle physics takes a most prominent place \cite{class}. For a thorough introduction into the
physical application of almost-commutative geometry we refer to \cite{schuck}.

\section{Constructing the Diagrams}

Our task will now be to construct an algorithm that finds the irreducible Krajewski diagrams. 
To begin with we will give the basic rules that follow from
the axioms of spectral triples and from the requirement of irreducibility. 
These rules will allow us to generate the diagrams and to perform all the necessary operations.

\subsection{Basic Rules}

\subparagraph{The diagram body:} The diagram body is the basis to construct diagrams from. It is
a quadratic array of a given size with circles indicating the intersection points or the rows and
the columns.  These circles have no  meaning on their own but are there to 
guide the eye of the reader. The size of the diagram body will be fixed in advance.

\begin{center}
\begin{tabular}{c}
\rxyz{0.7}{} \\ \\
$3 \times 3$ diagram body 
\end{tabular}
\end{center}

\subparagraph{The basic arrow:} The basic simple arrow connects two circles of the diagram body.
With the end points of an arrow we associate two numbers which are called the chirality. This term 
stems from noncommutative geometry and corresponds to the chirality of particle physics.
A simple arrow points from chirality $+1$ to chirality $-1$. 

\begin{center}
\begin{tabular}{c}
\rxy{
,(0,0)*\cir(0.7,0){}
,(5,0)*\cir(0.7,0){}
,(5,0);(0,0)**\dir{-}?(.6)*\dir{>}
,(0,-3)*{-1}
,(5,-3)*{+1}
} \\ \\
Simple Arrow
\end{tabular}
\end{center}

\subparagraph{Constructing a basic diagram:} To construct a basic diagram we put $n$ horizontal
arrows in an arbitrary way into the body. Only an antiparallel arrangement of arrows is forbidden.

\begin{center}
\begin{tabular}{c}
\rxy{
,(0,0)*\cir(0.7,0){}
,(5,0)*\cir(0.7,0){}
,(5,-0.3);(0,-0.3)**\dir{-}?(.6)*\dir{>}
,(0,0.3);(5,0.3)**\dir{-}?(.6)*\dir{>}
}\\ \\
Forbidden antiparallel arrows
\end{tabular}
\end{center}
Diagrams are sorted by their total number of arrows.
The empty diagram (i.e. the diagram body) is said to lie in level 0, the diagrams with one arrow are
said to lie in level 1 and so on. For $3 \times 3$ diagrams  some examples of possible diagrams are:
\begin{center}
\begin{tabular}{cccc}
\rxyz{0.5}{} \quad \quad & 
\rxyz{0.5}{
,(5,-5);(10,-5)**\dir{-}?(.6)*\dir{>}
} \quad \quad &
\rxyz{0.5}{
,(10,-5);(15,-5)**\dir{-}?(.6)*\dir{>}
,(15,-15);(5,-15)**\crv{(10,-13)}?(.6)*\dir{>}
} \quad  \quad &
\rxyz{0.5}{
,(5,-5.3);(10,-5.3)**\dir{-}?(.6)*\dir{>}
,(5,-4.7);(10,-4.7)**\dir{-}?(.6)*\dir{>} 
,(15,-10);(5,-10.1)**\crv{(10,-13)}?(.6)*\dir{>}
,(15,-15);(10,-15)**\dir{-}?(.6)*\dir{>}
,(5,-9.9);(10,-10)**\dir{-}?(.6)*\dir{>}
}
\\ \\
Level 0 & Level 1 & Level 2& Level 5 
\end{tabular}
\end{center}

\subparagraph{The Multiplicity Matrix:}
To every diagram a matrix, called the {\it multiplicity matrix} $M$, is associated. It is a direct translation
of the diagram by taking the values of the chiralities at each end of an arrow. These values are then
written into a square matrix of the same size as the diagram and at the same position as they appear in the diagram.
The multiplicity matrix is then this matrix plus its transposed as shown in the following example:

\begin{center}
\begin{tabular}{cc}
\rxyz{0.5}{
,(10,-5);(15,-5)**\dir{-}?(.6)*\dir{>}
,(15,-15);(5,-15)**\crv{(10,-13)}?(.6)*\dir{>}
,(20,-10);(25,-10)**\dir{-}?(1)*\dir{>} 
} 
&
\rxy{
,(10,-10)*{M = \pp{0&1&-1\\0&0&0\\-1&0&1} + \pp{0&0&-1\\1&0&0\\-1&0&1} = \pp{0&1&-2\\1&0&0\\-2&0&-2}}
} \\ \\
\end{tabular}

The multiplicity matrix of a diagram
\\[0.8cm]
\end{center}

\subparagraph{Clipping two arrows:} There are two ways to reduce a diagram. The first is by combining or {\it clipping} two arrows at a common point with common chirality or by building a corner again with arrow heads of common chirality. This creates a
multiple arrow.  Building
a corner is achieved by transposing one of the arrows (reflecting it on the main diagonal of the diagram)  which are to be clipped. 

\begin{center}
\begin{tabular}{ccc}
\rxyz{0.5}{
,(5,-5.3);(10,-5.3)**\dir{-}?(.6)*\dir{>}
,(5,-4.7);(10,-4.7)**\dir{-}?(.6)*\dir{>} 
,(20,-10);(25,-10)**\dir{-}?(1)*\dir{>} 
}&
\rxyz{0.5}{
,(5,-5)*\cir(0.2,0){}*\frm{*}
,(5,-5);(10,-5)**\dir2{-}?(0.6)*\dir2{>}
,(20,-15)*{,}
}&
\rxyz{0.5}{
,(10,-5)*\cir(0.2,0){}*\frm{*}
,(5,-5);(10,-5)**\dir2{-}?(0.6)*\dir2{>}
} \\
\rxy{
,(10,-5)*{M=\pp{4&-2&0 \\ -2&0&0 \\ 0&0&0 }}
,(20,-5);(25,-5)**\dir{-}?(1)*\dir{>} 
}&
\rxy{
,(10,-5)*{M=\pp{2&-2&0 \\ -2&0&0 \\ 0&0&0 }}
,(19,-8)*{,}
}&
\rxy{
,(10,-5)*{M=\pp{4&-1&0 \\ -1&0&0 \\ 0&0&0 }}
} \\ \\
\end{tabular}

Building a double arrow by clipping two parallel arrows 

\begin{tabular}{ccc}
\rxyz{0.5}{
,(5,-5);(10,-5)**\dir{-}?(.6)*\dir{>}
,(10,-10);(5,-10)**\dir{-}?(.6)*\dir{>}
,(20,-10);(25,-10)**\dir{-}?(1)*\dir{>} 
}&
\rxyz{0.5}{
,(10,-5)*\cir(0.2,0){}*\frm{*}
,(5,-5);(10,-5)**\dir{-}?(.6)*\dir{>}
,(10,-10);(10,-5)**\dir{-}?(.6)*\dir{>}
} \\ 
\rxy{
,(10,-5)*{M=\pp{2&-2&0 \\ -2&2&0 \\ 0&0&0 }}
,(20,-5);(25,-5)**\dir{-}?(1)*\dir{>} 
}&
\rxy{
,(10,-5)*{M=\pp{2&-1&0 \\ -1&2&0 \\ 0&0&0 }}
} \\ \\
\end{tabular}

Building a corner by clipping an arrow and a transposed arrow
\\[0.8cm]
\end{center}

The procedure to clip arrows extends in a straight forward way to more than two arrows. One 
only has to keep in mind to clip the arrows always at the same point in each connected component. Even two multiple arrows
can be combined at their clipping points. Clipping arrows will not change the level a diagram
belongs to.

\subparagraph{Erasing an arrow from a diagram:} The second way to reduce a diagram is by erasing an arrow.
Erasing an arrow from a diagram is
completely trivial and lowers the level of the diagram by one. A simple arrow will be deleted and an arrow which is clipped to another (multiple) arrow
will be deleted while leaving the element of the multiplicity matrix at clipping point unaltered:

\begin{center}
\begin{tabular}{cccc}
\rxyz{0.5}{
,(10,-5)*\cir(0.2,0){}*\frm{*}
,(5,-5);(10,-5)**\dir2{-}?(.6)*\dir{>}
,(10,-10);(10,-5)**\dir{-}?(.6)*\dir{>}
,(10,-15);(15,-15)**\dir{-}?(.6)*\dir{>}
,(20,-10);(25,-10)**\dir{-}?(1)*\dir{>} 
}&
\rxyz{0.5}{
,(10,-5)*\cir(0.2,0){}*\frm{*}
,(10,-10);(10,-5)**\dir{-}?(.6)*\dir{>}
,(5,-5);(10,-5)**\dir{-}?(.6)*\dir{>}
,(10,-15);(15,-15)**\dir{-}?(.6)*\dir{>}
,(20,-15)*{,}
}&
\rxyz{0.5}{
,(10,-5)*\cir(0.2,0){}*\frm{*}
,(5,-5);(10,-5)**\dir2{-}?(.6)*\dir{>}
,(10,-15);(15,-15)**\dir{-}?(.6)*\dir{>}
,(20,-15)*{,}
}& 
\rxyz{0.5}{
,(10,-5)*\cir(0.2,0){}*\frm{*}
,(5,-5);(10,-5)**\dir2{-}?(.6)*\dir{>}
,(10,-10);(10,-5)**\dir{-}?(.6)*\dir{>}
}\\ \\
\end{tabular}

All possibilities to erase one arrow from a diagram
\\[0.8cm]
\end{center}

\subparagraph{Irreducible Diagrams:} A diagram is said to be irreducible if it satisfies two conditions.
First, the determinant of the multiplicity matrix has to be non-zero. Second, reducing the diagram by clipping
or erasing arrows in all possible ways until one is left with the bare diagram body will on the way always produce diagrams with $\det M=0$.

\subparagraph{The Labels:}Every diagram carries a label indicating the reducibility. An
`i' stands for an irreducible diagram (the determinant of the multiplicity matrix $M$ is necessarily non-zero). Diagrams which carry 
an `r' can be reduced to a diagram equipped with an `i' or an `r' ($\det M$ is of no importance). The label `0' is carried by
diagrams where the determinant of the multiplicity matrix is zero and which cannot be reduced to diagrams
equipped with an `r' or an `i'. The labels are of internal use only and we usually do not write them explicitely. 

\subparagraph{Equivalence of Diagrams:} 
The physical theory following from a diagram is not
dependent on the order of the rows and columns or on a reversal of all arrows at once. Permuting the
rows and columns together or reversing the arrows does not change the absolute value of the determinant of the multiplicity matrix. 
Consequently diagrams
that differ only by permutations of columns and rows or by reversing the arrows will be regarded
as equivalent. All these diagrams build an equivalence class and only one
representative will be used for further operations. As an example the following three diagrams are 
equivalent: 

\begin{center}
\begin{tabular}{ccc}
\rxyz{0.5}{
,(5,-5)*\cir(0.2,0){}*\frm{*}
,(10,-5);(5,-5)**\dir{-}?(.6)*\dir{>}
,(5,-10);(5,-5)**\dir{-}?(.6)*\dir{>}
,(15,-15);(10,-15)**\dir{-}?(.6)*\dir{>}
,(20,-10)*{\simeq}
}&
\rxyz{0.5}{
,(15,-15)*\cir(0.2,0){}*\frm{*}
,(5,-15);(15,-15)**\crv{(10,-13)}?(.6)*\dir{>}
,(15,-5);(15,-15)**\crv{(13,-10)}?(.6)*\dir{>}
,(10,-10);(5,-10)**\dir{-}?(.6)*\dir{>}
,(20,-10)*{\simeq}
}&
\rxyz{0.5}{
,(5,-5)*\cir(0.2,0){}*\frm{*}
,(10,-5);(5,-5)**\dir{-}?(.4)*\dir{<}
,(5,-10);(5,-5)**\dir{-}?(.4)*\dir{<}
,(15,-15);(10,-15)**\dir{-}?(.4)*\dir{<}
}\\ \\
&Three equivalent diagrams&\\
\end{tabular}
\end{center}
With regard to several thousands of diagrams it becomes 
a very time consuming procedure to directly test all diagrams for possible equivalence. To compare two diagrams the set of arrows of one of the diagrams has to be permuted in all possible ways and reversed. All the  resulting diagrams
have to be compared to the other diagram.  Sorting the diagrams of one level into
equivalence classes by direct comparison is most uneconomical.

But there exist invariants of the diagrams. These
will provide us with a quicker way to decide whether two diagrams are inequivalent or not. Only if all
these invariants prove to be equal for two diagrams it becomes necessary to decide the question of equivalence by permuting and reversing the arrows:
 
\begin{description}
\item[1.] Compare directly the arrows of the two diagrams. If the arrows coincide
the diagrams are equal and thus equivalent.
\item[2.] Compare the absolute values of the determinants of the multiplicity matrices. If they do not coincide the diagrams are inequivalent.
\item[3.] Compare the absolute values of the traces of the multiplicity matrices. If they do not coincide the diagrams are inequivalent.
\item[4.] Compare the Chern character of the multiplicity matices. If they do not coincide the diagrams are inequivalent. This Chern character of the multiplicity matrix is defined as $\det (1+ M^2)$.
\item[5.] If the two diagrams are not inequivalent by virtue of points 1 to 4, permute and reverse the 
set of arrows and compare the resulting diagrams.
\end{description}
It is remarkable that the operations of point 1 to 4 are powerful enough to find out in most  
cases whether two diagrams are inequivalent. This reduces the computation time most considerably.
\label{equvalence}

\subparagraph{The Game:}
With the basic rules formulated above our combinatorial game is to find all the equivalence classes of
irreducible Krajewski diagrams for a given size of the diagram body and for a given maximal number of arrows.
This seemingly simple task will produce complex structures which we will call {\it diagram nets}. These nets fall into two categories, the main-nets and the clip-nets. We
will now present an algorithm to generate these nets and to find the embedded irreducible Krajewski diagrams.

\section{The Algorithm}
The algorithm to find the irreducible Krajewski diagrams can be
divided into four subalgorithms. Every subalgorithm stands independently but builds up on the
data produced by its predecessor. Summarised quickly the subalgorithms achieve the following:
The first subalgorithm, dubbed the ``main-net subalgorithm'' creates a diagram net consisting of all
equivalence classes of diagrams with up to $N$ arrows. The ``clip-net
subalgorithm'' creates a diagram net of equivalence classes from every element in the main-net, reducing each
diagram by clipping the contained arrows in all possible ways. The third subalgorithm which we call the ``label
subalgorithm'' checks whether the elements of a clip-net represent an irreducible diagram
and sets the  labels `i', `r' and `0' accordingly. Since the third subalgorithm can only see the elements
in a single clip-net the equivalence classes carrying an `i' or a `0' might still be reducible to an element
one level below by erasing an arrow. So the last subalgorithm, the  ``label correction subalgorithm'' 
checks these diagrams on their reducibility and changes their label to `r' if necessary. The output
of the whole algorithm are the diagrams labeled with an `i' representing irreducible Krajewski diagrams.

\subsection{The Main-Net Subalgorithm}
The ``main-net subalgorithm'' fills the diagram body with up to $N$ simple arrows. The maximal number
of arrows, i.e. the maximal level, is chosen before. The arrows are put into the diagram body
horizontally in every possible way, excluding only antiparallel arrows. Starting with the empty
diagram body level 1 is filled with one horizontal arrow. The resulting diagrams are checked if they are
equivalent and only one representative from each equivalence class is kept. In the same way every
following level is built from the representative diagrams of the equivalence classes by adding one arrow.
Each equivalence class is connected to its predecessors and its successors. Usually an equivalence
class has several predecessors. 

\begin{figure}[h!]
\begin{center}
\includegraphics[scale=0.9]{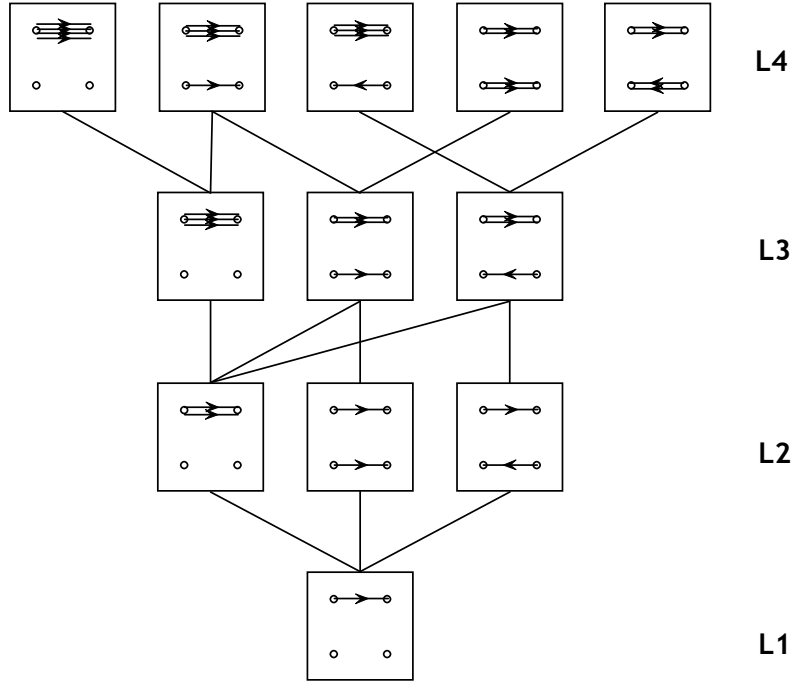}
\end{center}
\caption[label]{A main-net for a $2\times 2$ diagram with up to four arrows}
\label{exmain}
\end{figure}

\begin{figure}[f]
\begin{center}
\includegraphics[width=15cm]{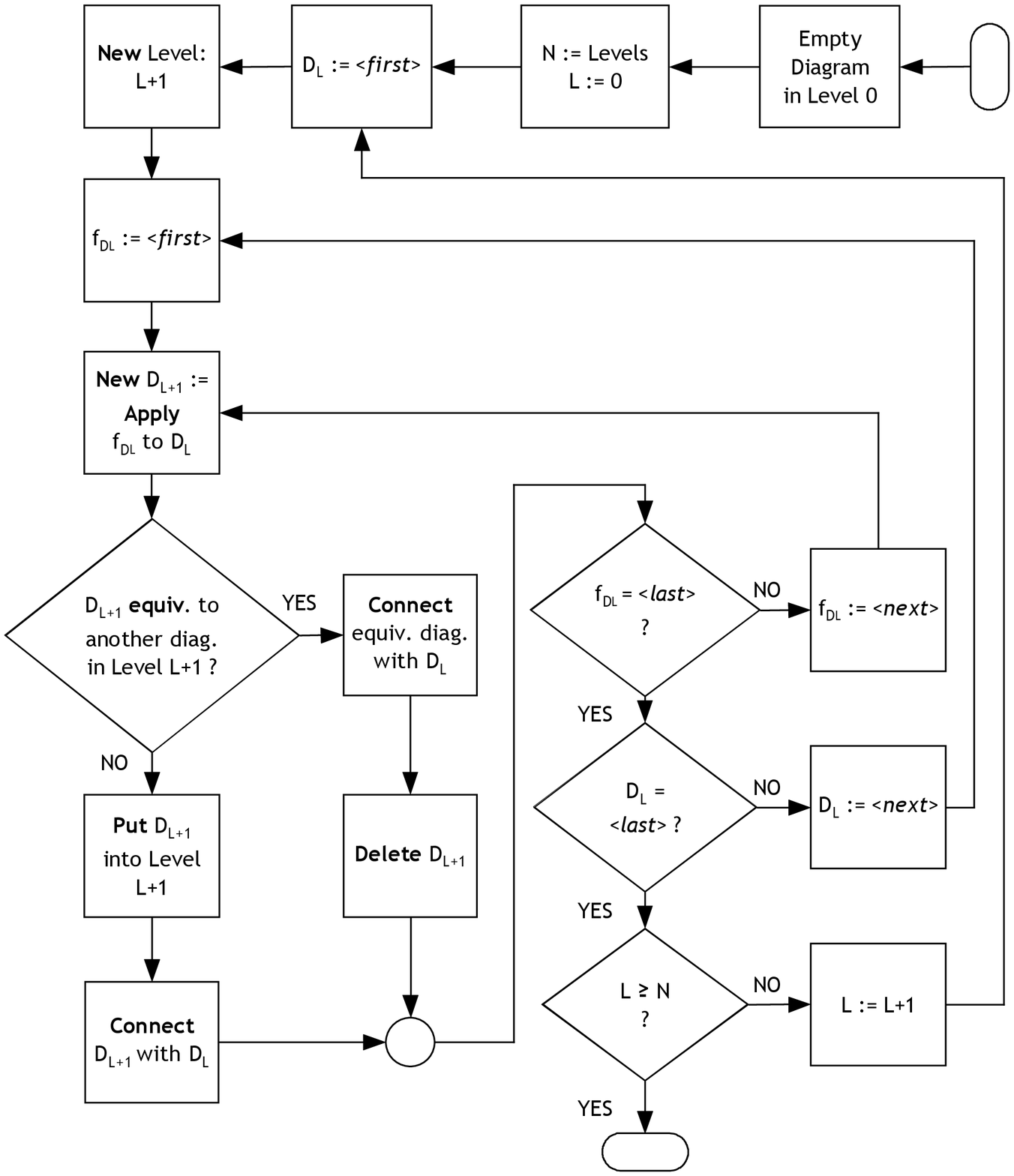}
\end{center}
\caption{The Main-Net Subalgorithm}
\label{maintree}
\end{figure}

The subalgorithm to generate the main-net is depicted in flow chart 1 in figure \ref{maintree}. Some explanations to read 
the flowchart are necessary:
\begin{description}
\item[1.] $N$ is the maximal number of arrows and is put in by hand. 
\item[2.] $L$ is the level index from which the diagrams are taken.
\item[3.] $D_L$ is an element of the ordered set that contains one representative of
each equivalence class on level $L$. This kind of set is a basic notion in our flow charts. The specific
order in the set is arbitrary, its only raison d'\^etre is  that the set can
be run through from the first to the last element. This notion of ordered
set has as its aim to unclutter the flow charts and to eliminate superfluous indices. 
\item[4.] $f_{D_L}$ is an element of the ordered set that contains all the operators which put one simple arrow
into the diagram $D_L$ in an allowed way. 
Each of these ordered sets is to be understood to belong to the specific diagram $D_L$ that is processed in the current loop.
\item[5.] $D_{L+1}$ is the diagram that is generated by putting in an arrow by virtue of the operator
 $f_{D_L}$. 
\item[6.] The equivalence of diagrams is checked making use of the definitions and invariants specified in 
section 2.1.
\end{description}

An example of a main-net for a $2\times 2$ diagram body with up to four arrows is given in figure \ref{exmain}.

\subsection{The Clip-Net Subalgorithm}
This algorithm creates a clip-net from each main-net diagram by clipping the arrows in all 
possible ways. The main-net diagram is said to lie in clip level 0. The diagrams where two arrows
have been  clipped are said to be in clip level 1, and so on. Each of the diagrams in the clip-nets
is again gathered in equivalence classes which are connected with regard to their predecessors and
successors.  Here a technicality from the noncommutative geometry comes in: Whenever a corner 
is built by clipping the diagram has to `forget' its predecessor. For a mathematical justification
of this rule we refer to \cite{class}.
The maximal number of clip levels is determined by
the number of arrows in the main-net diagram. If there are $L$ arrows in a diagram there are 
up to $L-1$ ways to clip these arrows.

\begin{figure}[h!]
\begin{center}
\includegraphics[width=12cm]{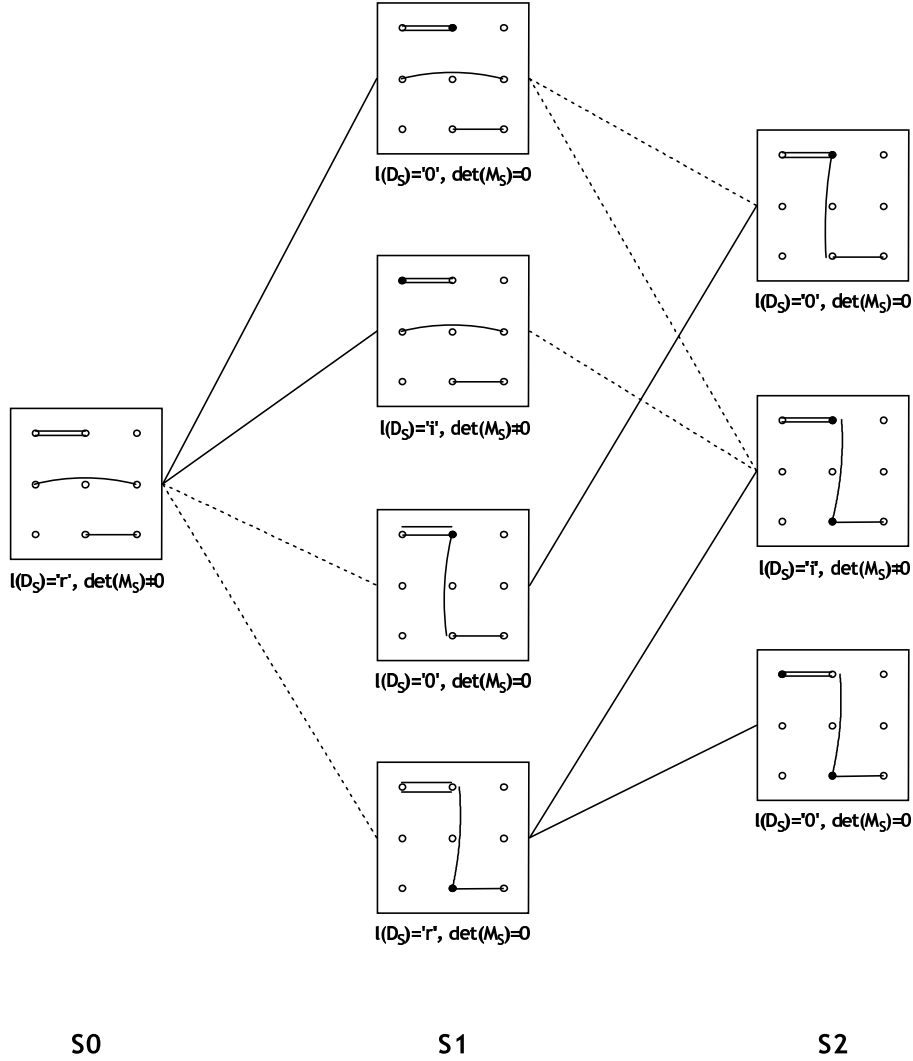}
\end{center}
\caption[label]{A simplified clip-net with 4 arrows and an example for the labelling}
\label{exsub}
\end{figure}

\begin{figure}[f]
\begin{center}
\includegraphics[width=14cm]{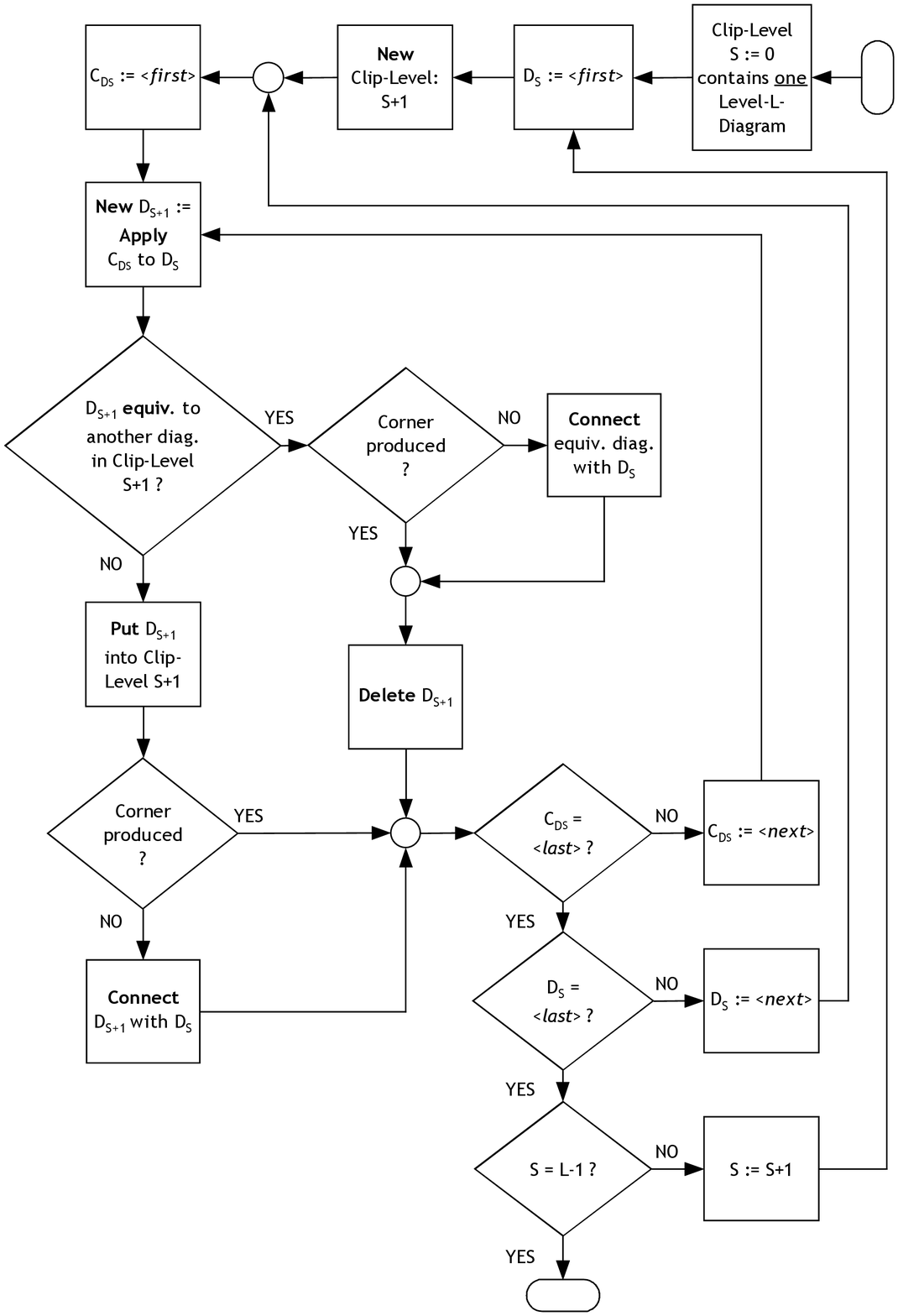}
\end{center}
\caption{The Clip-Net Subalgorithm}
\label{subtree}
\end{figure}

The subalgorithm to generate the clip-net is depicted in flow chart 2 in figure \ref{subtree}. The explanations to read 
the flowchart are:

\begin{description}
\item[1.] The representative of the equivalence class from the main-net is a member of the clip-net.
It is the only element on clip level 0.
\item[2.] $S$ is the clip level index. It can go up to $L-1$.
\item[3.] $D_S$ is an element of the ordered set of representatives of the equivalence classes of the diagrams
in clip-net $S$.
\item[4.] $C_{D_S}$ is an element of the ordered set of operators which clip pairs of arrows in the diagram  $D_S$, if possible. This set can be empty if it is impossible to clip any arrows.
\item[5.] $D_{S+1}$ is the diagram produced by applying $C_{D_S}$ to $D_S$, i.e. by clipping two arrows
in $D_S$.
\end{description}

A simple example of such a clip-net is given by figure \ref{exsub}. Here we have left out the arrows to visualise only the basic structure of a clip-net. The dashed lines represent the forgotten connections due to the building
of a corner.

\subsection{The Label Subalgorithm}
This subalgorithm assigns to each equivalence class in a clip-net one of the labels `i', `r' or
`0' (one should not forget that the elements of the main-net are included in the clip-net). For the labels `i' and `0' this assignment is provisional since it might be changed by
the ``label correction subalgorithm''. The subalgorithm starts with the highest clip level since here
the most reduced diagrams are situated. Here the labels are set to `i' or `0' depending
on the determinant of the multiplicity matrix of the diagram being non-zero or not. 
Then the algorithm descends one by one to 
clip level 0 setting the labels according to the determinant to `i' and `0' or setting the label
to `r' if the diagram is connected to a successor with label `i' or `r'.

\begin{figure}[f]
\begin{center}
\includegraphics[width=15cm]{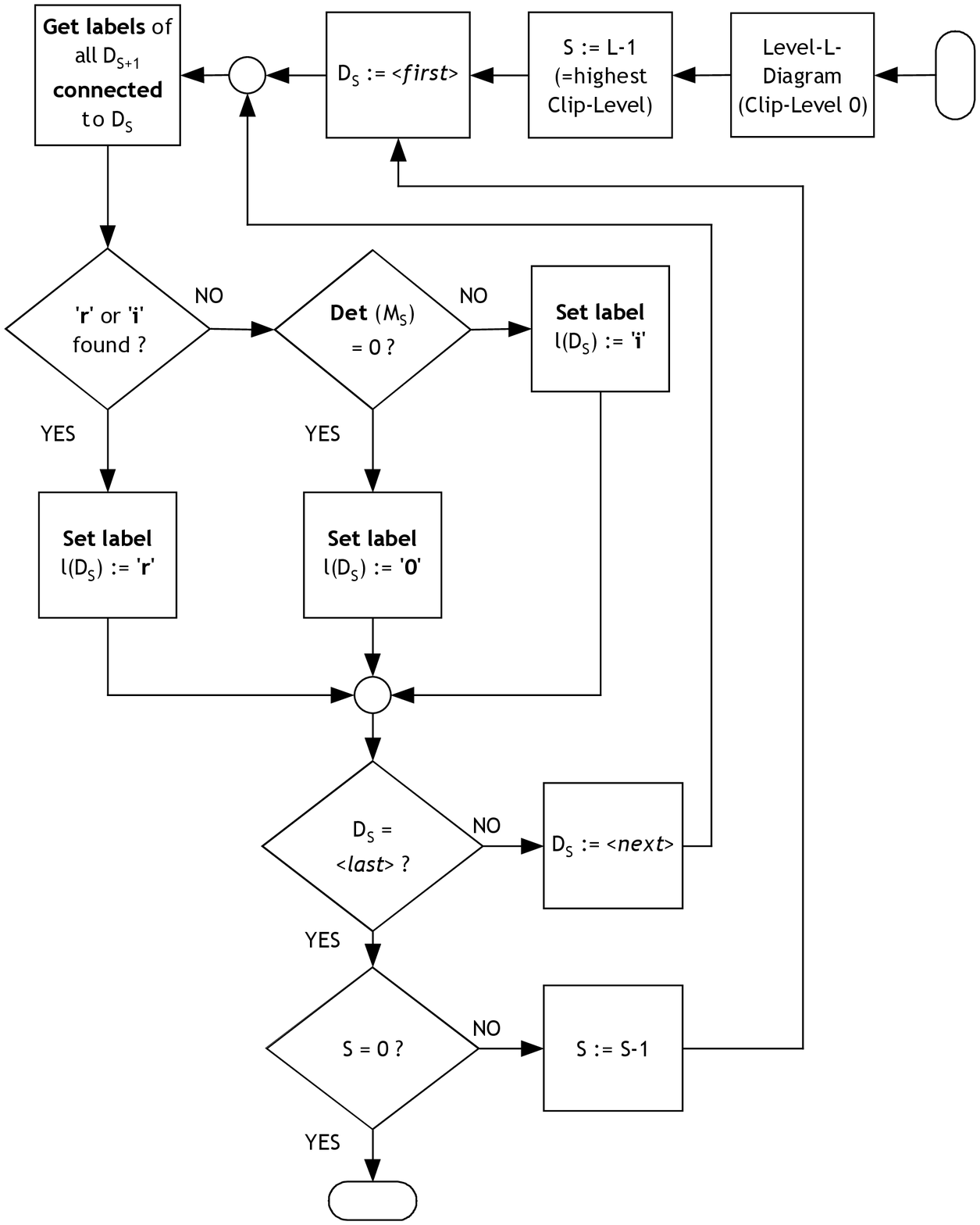}
\end{center}
\caption{The Label Subalgorithm}
\label{labsub}
\end{figure}

The subalgorithm to label the clip-net is depicted in flow chart 3 in figure \ref{labsub}. The explanations to read 
the flowchart are:

\begin{description}
\item[1.] If $S$ is the highest clip level the command ``Get label of $D_{S+1}$'' will always give
label `0' the since clip level $S+1$ does not exist and consequently there are no diagrams 
$D_{S+1}$.
\item[2.] $l(D_S)$ is the label of the element $D_S$ and takes values `i', `0' or `r'.
\item[3.] Det$(M_S)$ represents the determinant of the multiplicity matrix of $D_S$.
\end{description}

As an example for the labelling of a clip-net we also refer to figure \ref{exsub}. The labels and
the determinants are only examples.

\subsection{The Label Correction Subalgorithm}
\begin{figure}[f]
\begin{center}
\includegraphics[width=15cm]{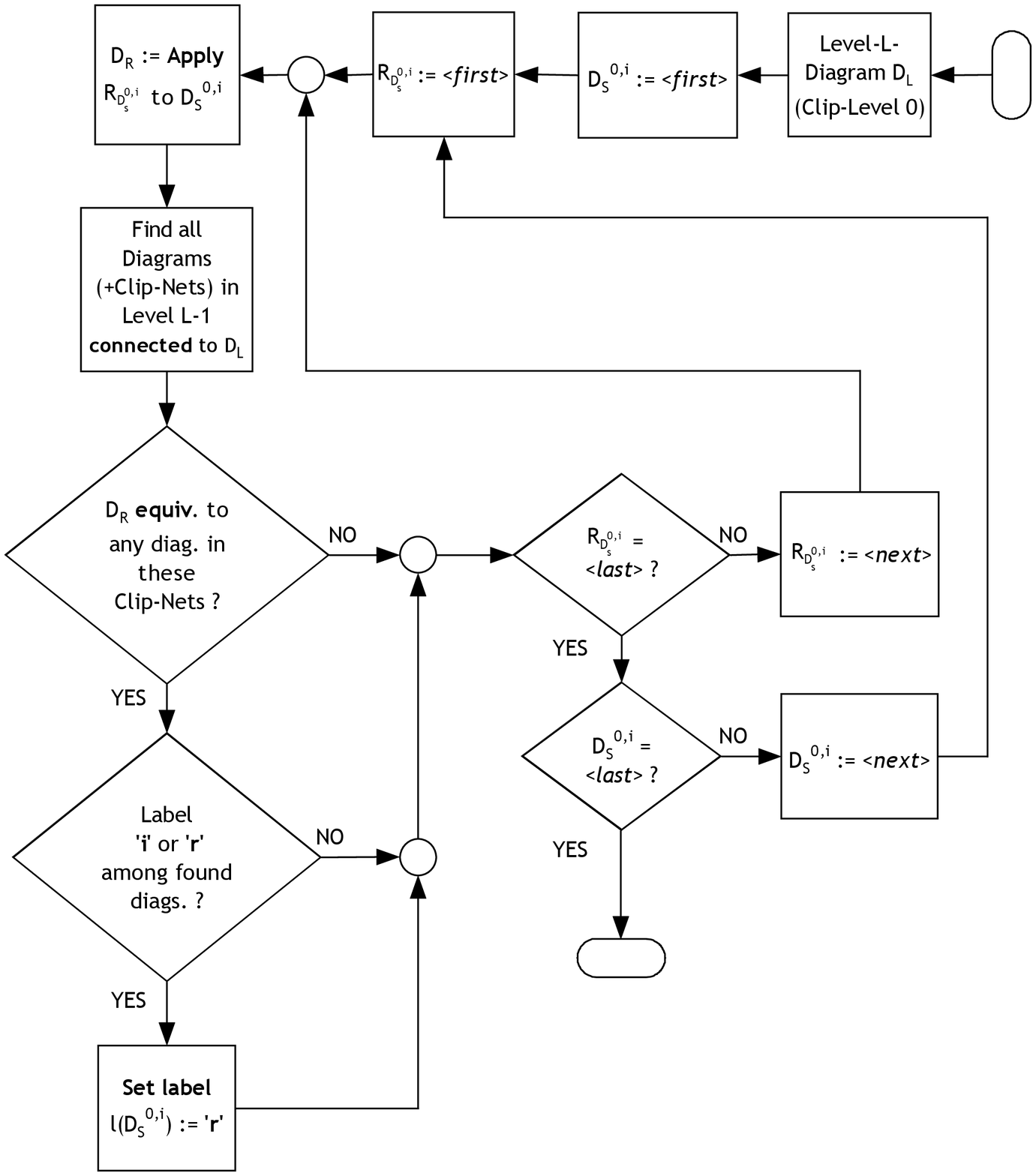}
\end{center}
\caption{The Label Correction Subalgorithm}
\label{labcor}
\end{figure}

Since it may be possible to reduce a diagram carrying label `i' or `0' to a diagram
with label `i' or `r' by erasing an arrow, it is necessary to correct these labels. It is evident that only
the clip-nets of the predecessors of the main-net diagram are of importance for relabelling. 
The subalgorithm to relabel the clip-net is depicted in flow chart 4 in figure \ref{labcor}. The explanations to read 
the flowchart are:

\begin{description}
\item[1.] $D_S^{0,i}$ is an element of the ordered set of {\it all} diagrams in the 
clip-net with label `i' or `0'. The relations among the elements in the clip-net are of no
importance since the diagrams are directly reduced to diagrams in clip-nets one level below.
\item[2.] $R_{D_S^{0,i}}$ is an element of the ordered set of operators that erase one arrow from
the diagram $D_S^{0,i}$.
\item[3.] $l(D_S^{0,i})$ is the label of the element $D_S^{0,i}$ and takes values `i', `0' or `r'.
\end{description}

\subsection{Assembling the Algorithm}

The main algorithm is composed of the four subalgorithms. Its input is the size of the diagram body
and the maximal number of arrows $N$. The assembly of the algorithm is now straight foreward:

\begin{description}
\item[Step 1:] Use the main-net subalgorithm to create the main-net up to level $N$.
\item[Step 2:] Create for each element in the main-net its clip-net with the clip-net
subalgorithm.
\item[Step 3:] Starting with level 1 label all clip-nets on this level with the label subalgorithm
and afterwards check the labels with the label correction algorithm. Repeat this procedure successively
for each level up to $N$.
\item[Step 4:] Print out all the elements with label `i'. These are the equivalence classes representing
irreducible Krajewski diagrams.
\end{description}

This algorithm runs very quickly on a recent PC with computation times of ca. 20 minutes for $4\times 4$ diagrams and five levels. But the complexity of the nets with their clip-nets grows rapidly with the size of
the diagram bodies and the maximal number of arrows. 

\section{Open Problems}

There are two major open problems. 
The first problem is the question how to choose the maximal number $N$ of arrows to be put into a diagram body.
We know that the total number of irreducible Krajewski diagrams is finite, see \cite{class}, but we do not
know the exact number of such diagrams. Up to now we adopted the simple conjecture that if no diagrams with
label `i' appear in two successive levels (say level 4 and level 5 for four algebras), we considered
the maximal number of irreducibles as reached. But there is still a possibility of ``islands'' of irreducible
Krajewski diagrams in higher levels.
Although we have checked all the elements of the diagram net  for $3\times 3$ diagram bodies and can be sure
that no more than three levels are necessary, already for $4\times 4$ diagram bodies this is no longer possible. We are
therefore looking for a criterion to determine the maximal level for a given size of the diagrams.

The second problem is the rapid increase of computational time and of memory required by the algorithm.
We quickly exceed the capabilities of an ordinary personal computer and it will probably be necessary
to use some sort of parallel computer or cluster to go beyond $5\times 5$ diagram bodies.

\section{Statement of the Result and Conclusions}

We presented an algorithm to find irreducible Krajewski diagrams. The rules given here quickly produce
very complex nets of diagrams in which the irreducibles are embedded. One of the most surprising features
of these nets is that for $3\times 3$ and $4\times 4$ diagram bodies only some twenty irreducible Krajewski
diagrams appear. A complete list of these diagrams can be found in \cite{class}. But already $5 \times 5$ diagram
bodies produce several hundreds of irreducible diagrams. The order of magnitude of the total  number of irreducibles seems to be
linked to the minimal number of arrows to  produce an irreducible diagram. $3 \times 3$ and $4 \times 4$ 
diagrams need two arrows whereas $5\times 5$ diagrams already need three arrows.  

The irreducible Krajewski diagrams are a vital ingredient for the classification of almost-commutative
geometries from the view point of a particle physicist. From this classification follows the prominent
position of the standard model within the Yang-Mills-Higgs theories compatible with noncommutative geometry.
To underline the virtue of the algorithm we would like to present the diagram corresponding to
the standard model of particle physics:

\begin{center}
\begin{tabular}{c}
\rxyz{0.7}{
,(10,-15)*\cir(0.4,0){}*\frm{*}
,(10,-5);(5,-5)**\dir{-}?(.6)*\dir{>}
,(10,-15);(5,-15)**\dir2{-}?(.6)*\dir2{>}
}
\\ \\ 
\end{tabular}

The Krajewsiki diagram of the standard model \\[0.8cm]
\end{center}

The actual program for the algorithm was written in $C^{++}$.

\vskip1cm
\noindent
{\bf Acknowledgements:} The authors would like to thank T. Sch\"ucker and B. Iochum for their advice and support. We gratefully acknowledge our fellowships of the Friedrich-Ebert-Stiftung.

\end{document}